\documentclass[a4paper,11pt]{article}
\usepackage{hyperref}
\usepackage[reqno,intlimits]{amsmath}
\usepackage{amsthm}
\usepackage{amssymb}
\usepackage{graphicx}
\usepackage[american]{babel}
\newcommand{\be}{\begin{equation}}
\newcommand{\ee}{\end{equation}}
\def\etal{\textit{et al.}\ }

\begin{document}
\bibliographystyle{utphys}
 \vspace*{1.8cm}
  \centerline{\bf HIGH ENERGY NEUTRINOS}
\vspace{1cm}
  \centerline{GIULIO AURIEMMA}
\vspace{1.4cm}
  \centerline{Universit\`a degli Studi della Basilicata, Potenza, Italy}
  \centerline{and}
  \centerline{INFN Sezione di Roma, Rome, Italy}
\vspace{0.5cm} \vspace{3cm}
% ----------------------------------------------------------------
\begin{abstract}
In this paper we compare the results of the MACRO detector at Gran
Sasso, that is the largest neutrino telescope operated before the
year 2k, with theoretical predictions of the neutrino emission
from some promising targets, such as blazars and GRB's. In
particular we propose a new statistical method, justified by the
maximum entropy principle, for assessing a model independent upper
limit to the differential flux of neutrinos inferred from the
measured up-ward going muon flux. This comparison confirms that
the acceptance of a detector of up-ward going muons should be of
the order of $1\;\mathrm{km}^2$, in order to challenge the present
theoretical estimates of possible neutrino production in both
galactic and extra-galactic objects.
\end{abstract}
\vspace{2.0cm}
% ----------------------------------------------------------------
\section{Introduction}\label{sect:intro}

Large detectors of high energy ($>1\;\mathrm{GeV}$) neutrinos,
like MACRO in the Gran Sasso Laboratory\cite{Ambrosio:2000yx}  or
SuperKamiokande \cite{Fukuda:1999pp} have been operated in the
last decade of the past century, but no extrasolar astrophysical
source has been detected. The occasion of this paper is related to
the decommissioning of the MACRO detector at Gran Sasso, that is
the largest neutrino telescope operated before the year 2k. The
proposal for the detector was accepted in November 1984
\cite{DeMarzo:1984cw}. Physics runs started in March 1989, as soon
as the first supermodule of the detector has been built
\cite{Calicchio:1988uw,Ahlen:1993pe} The full detector
\cite{Ahlen:1993kp} has been run from April, 1994 to December 31,
2000 when it has been decommissioned. A sample of a 1000 up-ward
going muons has allowed the measurement of the flux of geophysical
neutrinos collected up to September, 1999 \cite{Ambrosio:1998wu},
producing also interesting result \cite{Ambrosio:2001je}. The
purpose of this paper is to compare the upper limits obtained by
MACRO with current models of neutrino emission from astrophysical
objects. First in the following \S\ref{sect.detec} we will discuss
the physics of neutrino detection in detectors of up-ward going
muons. In particular in \S\ref{subsect:neutcs} we discuss the
extrapolation of the differential neutrino cross section to the
range of energies of interest for astrophysical neutrino
detection, in \S\ref{subsect:conv} the conversion of neutrinos
into muons in the rock and finally in \S\ref{subsect:uplim} we
give the rationale for a new statistical method for assessing a
model independent upper limit to the differential flux of
neutrinos inferred from the measured up-ward going muon flux. In
\S\ref{sect:results} we present some results obtained from the
MACRO survey of neutrino induced up-ward going muons. In
\S\ref{subsect:crab} we present the constraint to the neutrino
emission from the Crab nebula, in \S\ref{subsect:neutralino} that
for a possible neutralino decay in the Galactic center, in
\S\ref{subsect:blazars} the constraint to diffuse neutrino
background which can be inferred from the cumulative search of
coincidences with blazars and in \S\ref{subsect:grbs} we discuss
the same topic for the GRB deriving an upper limit to the diffuse
flux of neutrinos correlated with GRB . Finally in
\S\ref{sect:concl} we discuss the prospect for detection of
neutrinos in the next future by the planned detectors.

\section{Neutrino detection}\label{sect.detec}
\subsection{Extrapolation of neutrino cross
section}\label{subsect:neutcs} 
\begin{figure}[!ht]
\centering
\includegraphics[width=0.5\textwidth]{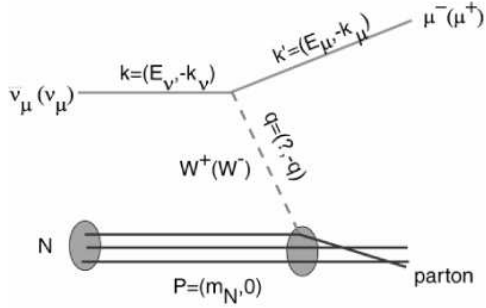}
\caption{\small The neutrino nucleon deep inelastic scattering in
the parton model.}\label{fig:parton-mod}
\end{figure}

The detection process for muon
neutrinos and anti neutrinos is the Charged Current (CC)
scattering illustrated in Fig.\ \ref{fig:parton-mod}
with double differential cross section
\begin{eqnarray}\label{eq:neut1a}
  &\displaystyle \frac{d^2\sigma_{\nu (\bar \nu)}}{dx\,dy} =
  \frac{G_F^2\,(s-m_N^2)}{2\pi}\,\frac{M_W^4}{(Q^2+M_W^2)^2}
  \nonumber \\ &\times
  \left\{\dfrac{y^2}{2}\,2 x
  F_1^{\nu(\bar \nu)}(x,Q^2)+\left(1-y\right)\,F_2^{\nu(\bar \nu)}(x,Q^2)+
  \pm\left(y-\dfrac{y^2}{2}\right)
  \,x F_3^{\nu(\bar \nu)}(x,Q^2)\right\}
\end{eqnarray}
where $y=1-E_\mu/E_\nu$ is the fraction of neutrino energy lost in
the Lab system, $x=Q^2/2m_N(E_\nu-E_\mu)$ the fraction of parton
momentum carried by the struck parton, and  $Q^2=x y (s-m_N^2)$
the square of the transferred four momentum. The last term is
positive for $\nu$'s and negative for $\bar \nu$'s. The three
functions $F_1,F_2$ and $F_3$ are the structure functions (SF's)
of  the nucleon. In the framework of the quark parton model (QPM)
the parton are identified with quarks, and the SF's can be
expressed in terms of the probability density functions (PDF) of
each quark types . The PDF's are the functions $q(x,Q^2)$ where
$q=u,d,s,c,b,t$ and the relative anti-quarks, which assign the
probability that a quark carries a momentum fraction $x$. In
particular we have for an isoscalar target
\begin{eqnarray}
F_2^{\nu N(\bar\nu N)} & = & x \left\{u+\bar u+ d+d\bar d+2 s+2c
+2b +2t\right\}\nonumber \\ x F_3 ^{\nu N(\bar\nu N)}& = & x
\left\{u-\bar u+ d-d\bar d\pm 2 s\mp 2c \pm 2b \mp
2t\right\}\nonumber \end{eqnarray} At very high energies the SF's
$F_2$ and $2\,x\,F_1$, are equal, but if the transverse momentum
of the $W$ is not negligible, they are related by the equation
\begin{equation}\label{eq:neut1z}
  2\,x\,F_1(x,Q^2)=F_2(x,Q^2)\;\frac{1+4 m_N^2 x^2/Q^2}{1+R(x,Q^2)}
\end{equation}
where $R(x,Q^2)=\sigma_L/\sigma_T$ is the ratio of the
longitudinal to transversal cross section of the $W$.
\begin{figure}[!ht]
\centering
\includegraphics[width=0.5\textwidth]{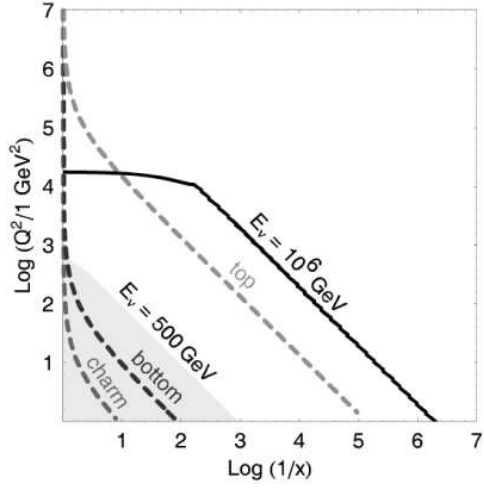}
\caption[h]{\small Physical integration domain over which Eq.\
(\ref{eq:neut1a}) must be integrated for a given energy of the
neutrino.}\label{fig:q2range}
\end{figure}

 The structure functions $F_2$ and $x F_3$ and the function $R$
 have been measured experimentally up
to $E_\nu\approx 500\;\mathrm{GeV}$ by the CCFR/NuTeV
collaboration
\cite{Yang:2001xc,Yang:2000ju,Bodek:2001hy,McDonald:2001vv}. To
calculate the neutrino cross sections at higher energies, as we
are interested in this paper, the SF's should be extrapolated. The
propagator term in Eq.\ (\ref{eq:neut1a})
${M_W^4}/{(Q^2+M_W^2)^2}\to 0$ for $Q^2\gg M_W^2$, therefore the
cross section at any energy will vanish for $Q^2\gtrsim
\max[xy(s-m_N^2),M_W^2]$. However the double differential cross
section of Eq\ (\ref{eq:neut1a}) must be integrated to obtain the
differential cross section $d\sigma/dy$ for a given $y$ in the
interval $ Q^2/y(s-m_N^2)\le x\le 1$. If the minimum detectable
energy for the muon in the detector is $E_{th}$ the lower
integration limit is $x_{min}\approx Q^2/2 m_N E_\nu$.
  In Fig.\ \ref{fig:q2range} we have
reported the area of the plane $\log_{10}(1/x),Q^2$ corresponding
to $E_\nu=10^6\;\mathrm{GeV}$ where Eq.\ (\ref{eq:neut1a}) must be
integrated, compared with the area where the SF's have been
measured.
\begin{figure}[!ht]
\centering
\includegraphics[width=0.5\textwidth]{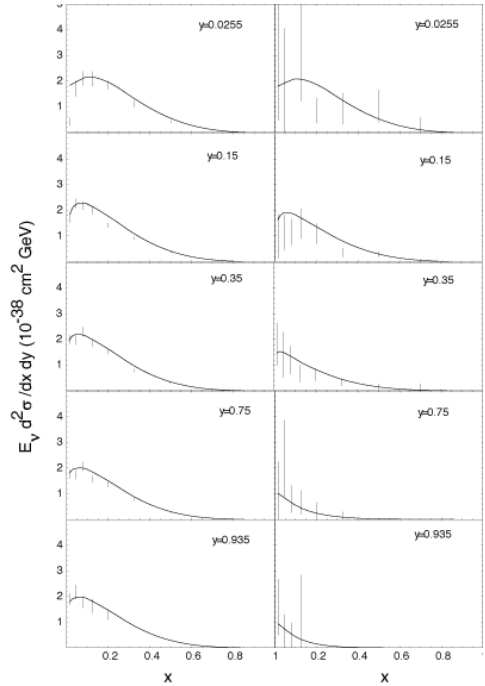}
\caption{\small Differential neutrino cross section at
$E_\nu=425\;\mathrm{GeV}$ measured by CCFR experiment. The fit is
the differential cross section obtained in this paper applying
CTEQ5 PDF's.}\label{fig:diffcs-ccfr-425}
\end{figure}

 In the same plot of Fig.\
\ref{fig:q2range}, we have reported the lines
$W^2=Q^2(1/x-1)=4\,m_h^2$ where $m_h$ are the masses of the heavy
quarks $c,b$ and $t$. Above these lines the production  of the
heavy quarks will give an important contribution to the cross
section. In fact it is well established that above $\approx
100\;\mathrm{GeV}$ the contribution of the quark production to the
total cross section is of the order of 20\% \cite{Bazarko:1995tt}.
From the plot of Fig.\ \ref{fig:q2range} we are forced to conclude
that the experimental knowledge of the physics underlying the
extrapolation of the cross section is, up to now, rather
incomplete. In the frame work of quantum cromodynamics (QCD), the
gauge field theory which describes the strong interactions of
colored quarks and gluons, the cross sections can be calculated
perturbatively in ascending order of the strong coupling constant
$\alpha_s$. In practice this corresponds to the solution of a set
of partial differential equations $\partial
F_i(x,Q^2)/\partial\ln(Q^2)$ (the DGLAP equations) at fixed $x$,
starting from the knowledge of the SF's $F_i(x,Q^2)$ with
$i=1,2,3$ at a given lower value $Q^2_0$ (see e.g. Sterman G.
\etal 1999 and references therein). This method has been
 checked, using the measured differential cross
sections. For example we report in Fig.\
\ref{fig:diffcs-ccfr-425} the fit to the measured differential
cross section obtained using PDF's calculated by the CTEQ5
collaboration \cite{Lai:1999wy}, with evolution equations
resulting from NLO analysis from a starting scale
$Q_0^2=1\;\mathrm{GeV}^2$. It is evident from this figure that the
agreement is reasonable but not perfect \cite{Kretzer:2001mb}.\par

\begin{figure}[!ht]
\centering
\includegraphics[width=0.6\textwidth]{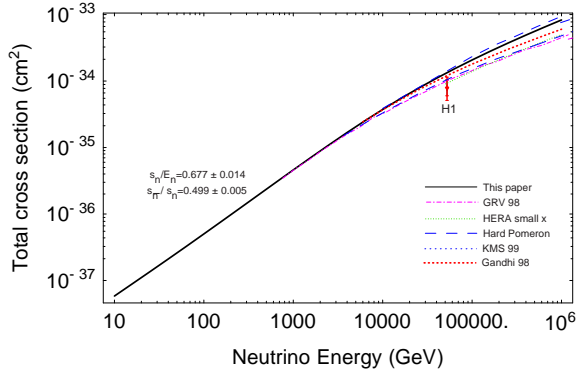}
\caption{\small Various extrapolations of the total cross section
for isoscalar target and equal mixture of $\nu_\mu$ and
$\bar\nu_\mu$. Solid line is the baseline estimate adopted in the
following of this paper (see text.). Finely dotted line is based
on the small $x$ structure functions measured at HERA, dash-dotted
on the dynamical parton model, which is practically coincident
with coarsely dotted line, dashed on the Regge theory for
$x<10^{-5}$ (see text). }\label{fig:csplot}
\end{figure}

Nevertheless the extrapolation at higher energies of the neutrino
cross section is still subjected to a non negligible uncertainty
because, as is shown in Fig.\ \ref{fig:q2range}, the domain of
integration at higher energies extends to smaller and smaller
values of $x$. Therefore, even if the evolution of SF's at fixed
$x$ to higher $Q^2$ can be calculated with the DGLAP equations,
still we lack of information about the SF's at low x values.
Various estimates, based on different physical assumption
\cite{Frichter:1995mx,Ralston:1996bb,Hill:1997iw,Gluck:1998js,Kwiecinski:1999bk},
as shown in Fig.\ \ref{fig:csplot} disagree by factors up to 2-4
for $E_\nu\gtrsim 10,000\;\mathrm{GeV}$. The experimental point
has been obtained by the H1 Collaboration \cite{Ahmed:1994fa}.
The higher values of the cross section is given by extrapolation
of the SF's for $x< 10^{-5}$ based on the Regge theory
\cite{Donnachie:1998gm,Berezinsky:2001hf}.

In the same Fig.\ \ref{fig:csplot} we have reported the cross
section of the reaction $e^- p\to \nu X$ by the H1 experiment at
HERA (Ahmed \etal 1994) which seems to be in better agreement with
the lower cross section estimate. In the rest of this paper we
will take as a baseline extrapolation of the cross section the
more conservative one, shown as a solid line in Fig.\
\ref{fig:csplot}, based on the CTEQ5 PDF's.
\subsection{Evaluation of conversion
probability}\label{subsect:conv} The probability
$P_{\mu\nu}(E_\nu)$ of detecting a muon with energy $\geq E_{th}$
in the detector, originated by a neutrino with energy $E_\nu$, is
given, in the continuous slowing down approximation, by the
integral \cite{Auriemma:1988ag}
\begin{equation}\label{eq:neut2a}
 P_{\mu\nu}(E_\nu)=\frac{1}{m_N}\, \int_0^{1-E_{th}/E_\nu}\,\left[a+b\,(1-y)E_\nu\right]^{-1}\,\int_0^1
  \frac{d^2\sigma_{\nu}}{dx\,dy}\,dy\,dx
\end{equation}
where $m_N$ is the mass of the nucleon, $a$ and $b$ are the
coefficient of the muon energy loss
$-{dE_\mu}/{dz}=a(E_\mu)+b(E_\mu)\,E_\mu$
 \cite{Groom:2000in}.
\begin{figure}[!ht]
\centering
\includegraphics[width=0.6\textwidth]{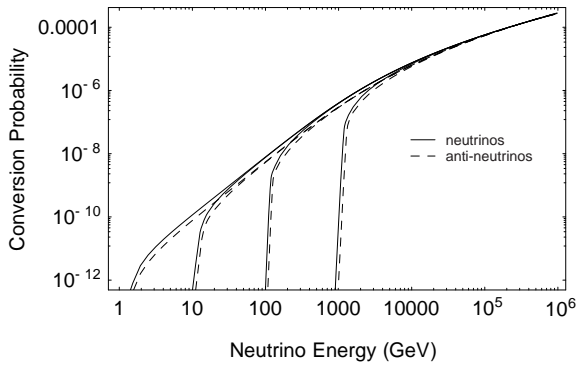}
\caption{\small Conversion probability for various $E_{th}$
values.}\label{fig:pmunu}
\end{figure}
\subsection{Model independent upper limits}\label{subsect:uplim}
The differential up-ward going muon flux in the detector will be $
d\Phi_\mu/dE_\nu=P_{\mu\nu}\, d\Phi_\nu/dE_\nu$ where
$d\Phi_\nu/dE_\nu$ is the spectral distribution of the neutrino
flux. However what is measured in an up-ward going muons neutrino
telescope, like MACRO, is only the integrated muon flux
\begin{equation}\label{eq:neut1d}
\Phi_\mu^{obs}(>E_{th})=\int_{E_{th}}^\infty\,P_{\mu\nu}\,
\frac{d\Phi_\nu}{dE_\nu}\,dE_\nu
\end{equation}
The conversion of the muon flux into a neutrino flux is subjected
to the ambiguities arising from the lack of knowledge of the shape
of the neutrino spectrum itself. The usual approach followed in
the literature \cite{Ambrosio:2000yx,Andres:2001ty} is to give the
upper limit to the neutrino fluxes assuming a spectrum
$d\Phi_\nu/dE_\nu\propto E_\nu^{-2}$. In this case the integrated
neutrino flux can be estimated using an average conversion
probability
\begin{equation}\label{eq:uplimz}
  \Phi_\nu(>E_{th})= \frac{\Phi_\mu^{obs}(>E_{th})}{\frac{1}{E_{th}}\,
  \int_{E_{th}}^\infty\,E_\nu^{-2}\,P_{\mu\nu}\,dE_\nu}
\end{equation}
We propose here to apply the maximum entropy principle
\cite{Jaynes:1957a,Jaynes:1957b} to infer the unknown neutrino
differential flux, with the less prior assumption on the spectral
shape. In fact, even if the maximum entropy principle has been
questioned for the lack of solid theoretical foundations, it gives
very significant results in several fields of applications of
spectral deconvolution \cite{Smith1983}. According to this
principle we assume that the probability distribution function of
the energy of the neutrino that could have produced a muon in the
MACRO detector will be uniform. In this case we have the ansatz
$P_{\mu\nu}\,d\Phi_\nu^{ME}/dE_\nu=\mathrm{const}$. Therefore,
normalizing to the integrated neutrino flux of Eq.\
(\ref{eq:uplimz}), we have derived the estimate of the upper limit
to the neutrino flux which will be shown in the following.
\section{Results of the MACRO survey}\label{sect:results}
\begin{figure}[!ht]
  \centering
  \includegraphics[width=0.6\textwidth]{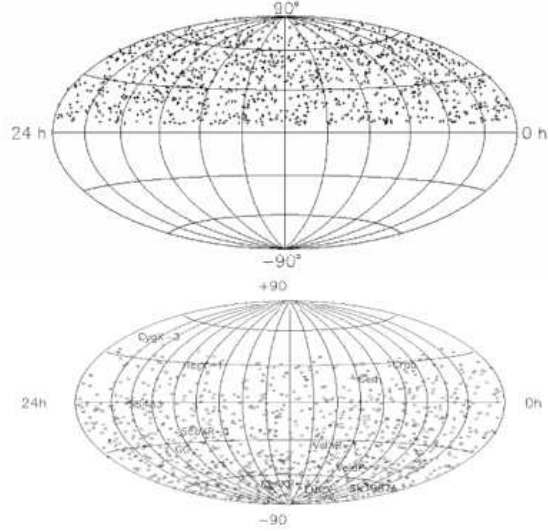}
\caption{\small Neutrino sky  after MACRO (lower panel). In the
upper panel is shown the complementary result of the AMANDA
survey.}\label{fig:amanda_Macro}
\end{figure}
\subsection{Neutrino emission from the Crab
nebula}\label{subsect:crab}

Gamma ray spectrum of the Crab Nebula extends up to 25 TeV. Above
$\approx 500\;\mathrm{GeV}$ the spectrum is a simple power law
with $\alpha\simeq 2.5$. The total $\gamma$-ray luminosity is
above $500\;\mathrm{GeV}$ $L_\gamma\simeq 1.5\times
10^{26}\;\mathrm{erg}\,\mathrm{cm}^{-2}\,\mathrm{s}^{-1}$. The
detection of neutrino emission from the Crab could support the
possibility that this high energy emission is due to hadronic
interactions. Bednarek \& Protheroe (1997) have analyzed the
consequences of acceleration of heavy nuclei in the pulsar
magnetosphere as a possible mechanism of energetic radiation from
the Crab Nebula. The MACRO limit for the Crab Nebula is
\begin{equation}\label{eq:craba}
  \Phi_\mu^{Crab}(>1.2\;\mathrm{GeV})\le
2.5\times10^{-14}\,\mathrm{cm}^{-2}\,\mathrm{s}^{-1}\quad
(90\%\;\mathrm{C.L.})
\end{equation}
\begin{figure}[!ht]
  \centering
 \includegraphics[width=0.6\textwidth]{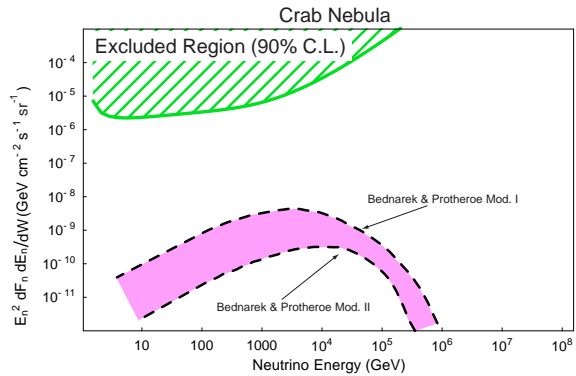}
  \caption{\small MACRO Upper Limit (90\% C.L.) for neutrino emission
from the Crab Nebula}\label{fig:crablim}
\end{figure}
In Fig.\ \ref{fig:crablim} we compare the limit on the neutrino
flux, with two extreme models. The present emission is determined
by the trapped high energy proton's density in the nebula.
Therefore the present emission is mainly due to acceleration in
the early times after supernova explosion. Magnetic dipole losses
determine the pulsar period at present, but the initial period is
determined largely by gravitational losses and it has been
probably shorter than 10 ms. Hence the two models consider two
initial periods: 5 ms, and 10 ms\cite{Bednarek:1997cn}.
\subsection{Neutralino annihilation in the Galactic
Center}\label{subsect:neutralino} 
\begin{figure}[!hb]
  \centering
  \includegraphics[width=0.55\textwidth]{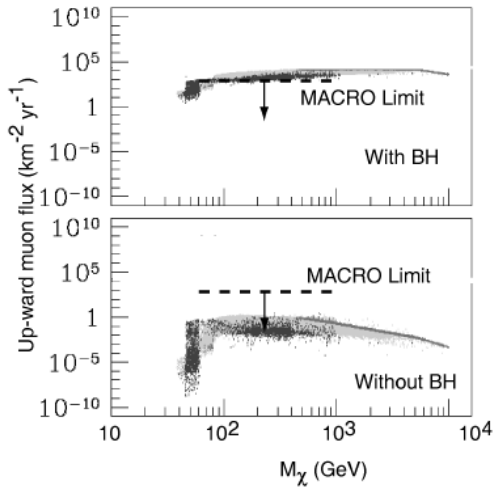}
\caption{\small Predicted up-ward going muon flux from Galactic
Center. The possible enhancement due to the redistribution of
neutralinos is shown in the upper panel. }\label{fig:GC_annih}
\end{figure}

Intense emission from the
Galactic center is predicted if cold dark matter is present there,
as in current models of the dark galactic halo
\cite{Berezinsky:1994wv,Tsiklauri:1998np,Gondolo:1999ef}. In
particular it could be expected, according to Gondolo \& Silk
(1999) that a massive black hole at the galactic center could
redistribute the WIMP's into a cusp. The effect of this
redistribution would be a strong enhancements of the
self-annihilation rate of neutralinos. In Fig.\ \ref{fig:GC_annih}
we report the predicted up-ward going muon flux in the two cases.
After the complete run MACRO \cite{Ambrosio:2000yx} the 90\% C.L.
upper limits on up-ward going muons from the Galactic Center is
$\Phi_\mu\le
0.34\times10^{-14}\,\mathrm{cm}^{-2}\,\mathrm{s}^{-1}$. This new
limit is reported in Fig.\ \ref{fig:GC_annih}. It is clear from
this figure that the predicted enhancement has not been observed.
One possible explanation of this negative result has been given
from a subsequent more careful evaluation, in a recent paper, of
the dynamics of the cusp formation \cite{Ullio:2001fb}. The MACRO
negative observation supports the possibility that the accreting
matter could not spiral fast enough by dynamical friction only.
Thus within a Hubble time only a mild enhancement could have taken
place.

\subsection{Diffuse emission from Blazars}\label{subsect:blazars}
\begin{figure}[!ht]
  \centering
 \includegraphics[width=0.6\textwidth]{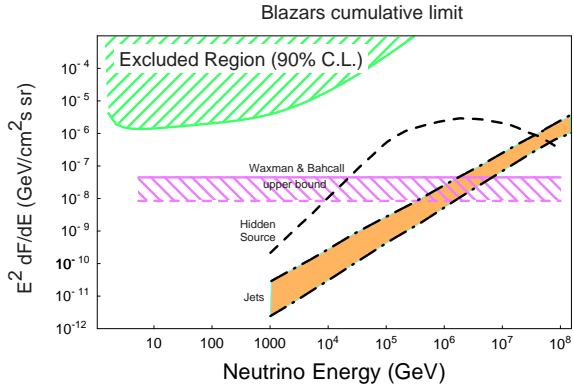}
  \caption{\small MACRO Upper Limit (90\% C.L.) for the diffuse neutrino emission
from blazars}\label{fig:blazarlim}
\end{figure}

The second EGRET catalog of high-energy $\gamma$-ray sources
\cite{Thompson:1995} contains 40 high confidence identification of
AGN and all appear to be blazars. This is the main reason for
which several authors \cite{Stecker:1996th,Protheroe:1996uu} have
proposed this type of object as potential powerful sources of
high energy neutrinos. The proton blazar model
\cite{Nellen:1993dw} both protons and electrons are accelerated
and protons interact with synchrotron radiation produced by
electrons photoproducing pions that decay into $\gamma$-rays and
neutrinos.

The up-ward going muons distribution, shown in Fig.\
\ref{fig:amanda_Macro} has been searched for angular coincidences
in a 3$^\circ$ cone with each of the 181 blazars, listed in a
recent catalogue \cite{Padovani1995}, with a declination
$\delta\le 40^\circ$. No significant excess has been found for any
individual source, with a cumulative average upper limit to the
muon flux $\Phi_\mu(\ge 1.2 \mathrm{GeV})\le5.44\times
10^{-16}\;\mathrm{cm}^{-2}\,\mathrm{s}^{-1}$. A bias in favour of
northern declinations is clearly present in the sample. In fact
178 out of the 233 objects listed in the catalog, have $\delta >
0^\circ$ and there are no known BL Lacs with $\delta < -57^\circ$.
We can estimate, very roughly, that the MACRO survey has covered
the declination band $-57^\circ\le \delta\le 40^\circ$,
corresponding to $\approx 3.2\,\pi\;\mathrm{sr}$ of sky. From this
figure we can estimate the U.L. to the diffuse up-ward going muons
from blazars
\begin{equation}\label{eq:blazarsa}
  \frac{d\Phi_\mu^{BLac}}{d\Omega}\le 0.98\times
  10^{-14}\;\mathrm{cm}^{-2}\,\mathrm{s}^{-1}\,\mathrm{sr}^{-1}\quad
  (90\%\;\mathrm{C.L.})
\end{equation}
Applying the maximum entropy method we obtain the upper limit to
the diffuse neutrino flux reported in Fig.\ \ref{fig:blazarlim}.
In the same plot we have reported the model independent upper
bound that can be derived if the UHE cosmic rays are also
accelerated in the same source \cite{Waxman:1998yy}. The higher
prediction is obtained, presuming a blazar's luminosity
cosmological evolution following that of QSO's, which may be
described as $f_{QSO}=(1+z)^3$ at redshift $z<1.9$ and constant
above. The lower prediction is for no evolution of the blazar's
density.
\subsection{Search for neutrinos from GRB's}\label{subsect:grbs} Several mechanisms for production of
intense HE and UHE neutrino and $\gamma$ burst associated with the
main pulse, in the $0.1\;\mathrm{MeV}$ have been proposed in the
literature
\cite{Vietri:1998nm,Waxman:1998tn,Guetta:2001cd,Meszaros:2001ms}.
It is somewhat surprising however that a phenomenon that was
discovered in the keV-MeV region could be the herald of big
activity in the multi-TeraVolt region.

\begin{figure}[!ht]
  \centering
  \includegraphics[width=0.6\textwidth]{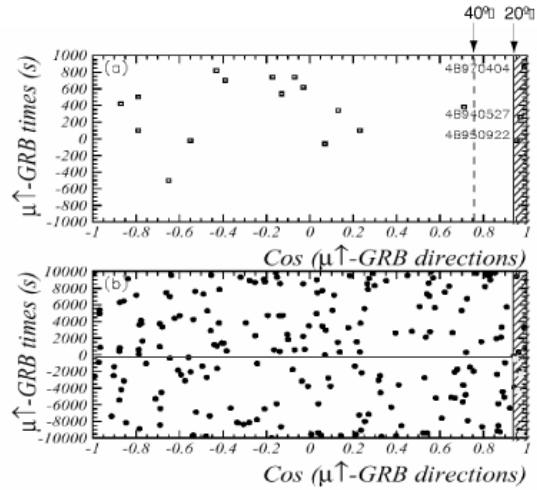}
  \caption{\small Coincidences of up-ward going muons in MACRO with the 2527 gamma ray
bursts from 21 Apr. 1991 to 5 Oct. 1999 (see text)
.}\label{fig:MACRO_GRB2}
\end{figure}

We have searched the catalogues BATSE 3B \& 4B
\cite{Paciesas:1999tp} containing 2527 gamma ray bursts from 21
Apr. 1991 to 5 Oct. 1999 \cite{Ambrosio:2000yx}. They overlap in
time with 1085 upward-going muons collected by MACRO during this
period. In Fig.\ \ref{fig:MACRO_GRB2} it is shown a plot of
angular distance among the GRB and up-ward going muons in MACRO
as a function of arrival time difference.
We observe that if neutrino are massive, as suggested by
observation of neutrino oscillations, the neutrino burst from
cosmological GRB will be delayed respect to the observation of the
$\gamma$ rays observed by the GRB itself. The time delay will be
\begin{equation}\label{eq:neut7a}
  \delta t=\frac{1-\beta}{\beta}(t_z-t_0)
\end{equation}
where $\beta=\sqrt{E_\nu^2-m^2_\nu c^4}/E_\nu$. For $E_\nu\gg
m_\nu c^2$ we can put ${(1-\beta)}/{\beta}=1/2 (E_\nu/m_\nu
c^2)^{-2}+\mathcal{O}(E_\nu/m_\nu c^2)^{-4}$.
\begin{figure}[!ht]
  \centering
  \includegraphics[width=0.6\textwidth]{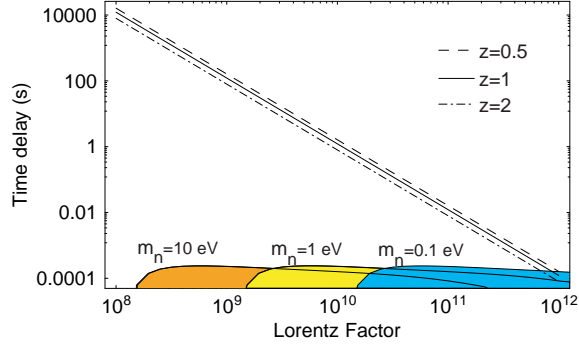}
  \caption{\small Time delay of the neutrino burst from the main
  pulse as a function of energies for different neutrino masses.}\label{fig:timedelay}
\end{figure}
In Fig.\ \ref{fig:timedelay} we show the expected delay of the
massive neutrino signal, for GRB in the range $z=0.5-2$ as a
function of the Lorentz factor of the neutrino. In the same figure
we have reported the normalized Lorentz factor distribution (for a
source spectrum $dN_\nu/dE_\nu\propto E_\nu^{-2}$) of neutrinos
that produce upward going muons in the MACRO detector. From this
figure we see that a maximum delay of $\approx 1000\;\mathrm{s}$
could be expected for very massive neutrinos
($m_\nu=10\;\mathrm{eV}$).
\begin{figure}[!ht]
  \centering
 \includegraphics[width=0.6\textwidth]{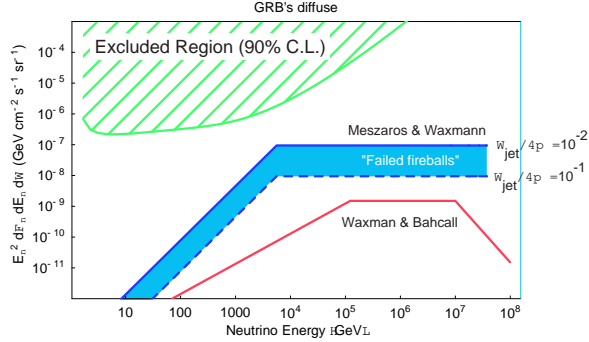}
  \caption{\small MACRO Upper Limit (90\% C.L.) for the diffuse neutrino emission
from GRB's.}\label{fig:grblim}
\end{figure}

Therefore the time window to be searched for should to be expanded
up to $\approx 1000$ s after the detection of the GRB for the muon
to be detected in MACRO. However according to the current ideas on
the production of the GRB explosion, it should not be possible to
have the neutrino emission long before the GRB. Therefore we can
assume that the coincidences between muons detected in the time
window preceding the GRB should be very likely accidental. From
the lower panel of Fig.\ \ref{fig:MACRO_GRB2} we estimate that the
accidental coincidences among up-going muons and GRB with
$\Delta\theta\le 20^\circ$ will be $0.4$ in the overlap period.
From the upper panel of the same figure we could candidate two
events as possible true coincidences, one event after 39.4
s from 4B950922  at an angular distance of 17.6$^\circ$ and
another very horizontal event in coincidence with the 4B940527
inside 280 s at 14.9$^\circ$. However the cumulative Poissonian
probability that those two events are accidental is $7.9\times
10^{-3}$. We assume then that those events are expected
accidental, and calculate that the 90\% confidence interval for the
unobserved true coincident upward-going muons
\cite{Feldman:1998qc} will be $\le 5.41$. The corresponding upper
limit to the muon fluence from the average GRB is $ \Phi_\mu(\ge
1.2\;\mathrm{GeV})\le 0.79\times 10^{-9}\;\mathrm{cm}^{-2}\;(90\%
\mathrm{C.L.})$ for neutrino masses $m_\nu\le 10\;\mathrm{eV}$.
From this cumulative limit we can derive an upper limit to the
diffuse neutrino background from GRB's, taking into account the
fact that the MACRO live time in the period from 21 Apr. 1991 to 5
Oct. 1999 has been $4.62\;\mathrm{y}$, that neutrinos could have
been detected for GRB's with declination $\delta\le 40^\circ$
(corresponding to $3.53\,\pi\;\mathrm{sr}$ of sky) and that the
exposure factor for BATSE in this declination range
\cite{Paciesas:1999tp} is $55\%$ we can convert the upper limit
for the average burst to a diffuse background
\begin{equation}\label{eq:neut2b}
  \frac{d\Phi^{GRB}_\mu}{d\Omega}\le 2.25\times
  10^{-15}\;\mathrm{cm}^{-2}\,\mathrm{s}^{-1}\,\mathrm{sr}^{-1}\quad
  (90\%\;\mathrm{C.L.})
\end{equation}
We obtain also in this case the upper limit to the diffuse
neutrino emission from GRB's reported in Fig.\ \ref{fig:grblim}.
In the same figure we have reported the prediction of diffuse
neutrino background produced by a cosmological distribution of
GRB's \cite{Waxman:1998yy}. Also we have reported the flux
expected from the intriguing possibility that before the jet could
emerge from the dense stellar progenitor of the GRB, the
accelerated protons could interact with the thermal X-ray photon
leading to neutrinos with energy $\gtrsim 5\;\mathrm{TeV}$
\cite{Meszaros:2001ms}. In this model the intensity of diffuse
neutrino background depends from the amount of collimation of the
jet $\Omega_{jet}/4\pi$, because the $\simeq 1000$ GRB's observed
per year, are only a small fraction of the actual explosions.
\section{Conclusions}\label{sect:concl}
\begin{figure}[!ht]
\centering
   \includegraphics[width=0.5\textwidth]{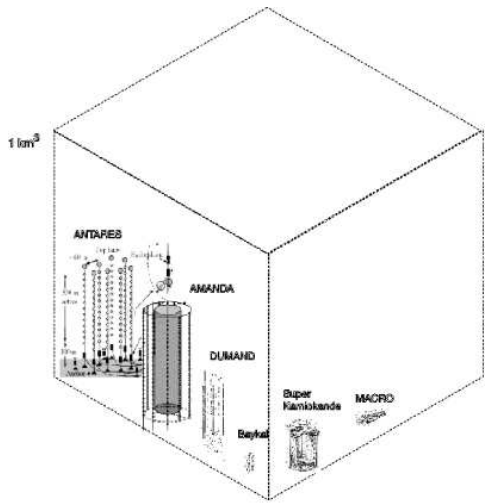}
\caption{\small A gallery of past, present and future HE neutrino
 telescopes.}\label{fig:detec_comp}
\end{figure}
In Fig.\ \ref{fig:detec_comp} we show a gallery of past, present
and future HE neutrino telescopes, compared with a massive
detector with 1 km$^3$ of volume. From this figure we have a
direct impression of the smallness of the detectors which were
operated in the past century, compared with the ones under
construction. It is also worth noticing that the preliminary
results obtained by Amanda, from April to October, 1997 which are
reported in the upper panel of Fig.\ \ref{fig:amanda_Macro},
include 1097 up-ward going muons. A number comparable to the
number of events detected by MACRO in $\approx 4.6$ live years.
This shows directly that the acceptance of the AMANDA-II detector
is already more then a factor 10 greater then the acceptance of
MACRO.

In the previous \S\ref{sect:results} we have practically shown how
far was MACRO from effectively testing theoretical production
models. From Fig\ \ref{fig:crablim} it can be crudely estimated
that an acceptance $\approx 10^{4}\;\mathrm{MACRO}\simeq
1\;\mathrm{km}^2$ will be needed to challenge possibly the
predictions for galactic neutrino sources. The situation could be
slightly better for the statistical detection of emission from
selected classes of objects, like for example the association of
up-ward going muons with blazars, where a possible detection seems
to be attainable with an acceptance $\approx 0.1\;\mathrm{km}^2$.

\bibliography{vulc01,waxman,auriemmag,halzen}
% \bibliography{auriemmag}        %or whatever your .bib file is
%\bibliographystyle{abbrvdin}   %if you use utphys.bst
\end{document}